\documentclass[aps,prl,amsmath,twocolumn]{revtex4}
\usepackage{bm}
\usepackage{graphicx}

\def \bu{{\bf u}}
\def \ba{{\bf a}}
\def \bO{{\bf O}}

\def \bGamma{\boldsymbol{\Gamma}}
\def \bsigma{\boldsymbol{\sigma}}

\def \br{{\bf r}}

\def \bQ{{\bf Q}}
\def \be{{\bf e}}
\def \bx{{\bf x}}
\def \bl{{\bf  l}}

\def \bR{{\bf R}}

\begin{document}

\title{Lagrangian stochastic integrals of motion in isotropic random flows }

\author{V.A. Sirota$^{1}$, A.S. Il'yn$^{1,2,}\footnote{%
Corresponding author: asil72@mail.ru}$, A.V. Kopyev$^1$, K.P.
Zybin$^{1,2}$}
\affiliation{$^1$ P.N.Lebedev Physical Institute of RAS, 119991, Leninskij pr.53, Moscow,
Russia \\
$^2$ National Research University Higher School of Economics, 101000,
Myasnitskaya 20, Moscow, Russia
}


\begin{abstract}

A set of exact integrals of motion is found for systems driven by homogenous isotropic stochastic flow.
 The
integrals of motion describe the evolution of (hyper-)surfaces of different dimensions transported
by the flow, and can be expressed in terms of local surface densities. The
expression for the integrals is universal: it represents general geometric properties and does not
depend on the statistics of the specific flow.
\end{abstract}

\maketitle

Stochastic systems with multiplicative noise play an important role in fundamental physics as well as in numerous physical, chemical and biological {applications~\cite{Zeld-interm, Vankampen, Pavliotis, Sethna}.} These are numerous problems related to non-equilibrium thermo- and hydrodynamics; in particular, systems with random flows including climate prediction~\cite{climate-1, climate-2}, transport of different kinds of admixtures in biology~\cite{bio, bio-new} and chemistry~\cite{chem, chem-new}, and magnetohydrodynamics~\cite{Rincon, apj, mnras} represent the simplest and bright example of a multiplicative system.
 Theoretical description of these problems is far from completion; it
faces essential difficulties caused by the fact that the central limit theorem is not valid for the
cases, and the wings of the distribution may affect the results crucially (cf. large deviations theory~\cite{Dembo, Touchette, Nick}).

In the  absence of complete solution, it is important to find some reference points, to reveal
general properties of the system. Conservation laws is the most informative example of such
properties. Ten years ago, one exact integral of motion was found in \cite{FalkFrishman}  for a couple
of particles driven by a random flow with stochastic homogeneous and isotropic
smooth divergentless velocity field.  Up to now, it has been the only known conserved quantity.

In this paper we derive the set of exact integrals of motion for arbitrary smooth homogenous
isotropic random flows in a space of arbitrary dimension; incompressibility is not needed.
 We also propose  a geometric interpretation for the integrals.
In the  case of incompressible flow, one of the integrals is reduced to that found in \cite{FalkFrishman}.

The result of \cite{FalkFrishman} was based on the work \cite{Zeld}; the integral of motion was
first found to be an asymptotic limit as $t\to \infty $ in a stationary flow. In
\cite{FalkFrishman} it was shown to be an exact time invariant.  Quite similarly, the integrals
discussed below were first found in \cite{PREL23} as a long-time limit for stationary flows. In
this paper it is proven that they are valid for any time, and for any type of flows including
nonstationary ones. However, despite the similarity of the two pairs of discoveries, the methods
were quite different in all these four cases.

The integrals of motion have the same form independently of the specific statistics of the flow.
Their existence turns out to be a consequence of
 very general property, purely geometrical, not even dynamical, feature of isotropic stochastic
configurations.
More specifically, it concerns statistics and geometry of random lines and surfaces frozen in the
flow.

The origin of these integrals  is intermittency. Actually, regardless of whether the velocity field
is intermittent~\cite{Toschi, gurevich} or not~\cite{Kraich-pass}, passive fields trapped by the flow always demonstrate intermittent
behavior. So do material lines and surfaces transported by the flow~\cite{FGV, Qi, CompLE}. Consider one selected
material line or plane, and follow its evolution in the flow; for turbulence, the illustration of
what happens to the line can be seen in \cite{Bentkamp}. In a random flow, there are always regions
of the material line that are very stretched, and the regions that are strongly contracted.
Let us mark the line/plane by some paint with homogenous initial linear/surface density and trace
the changes of the density along the deformed surface as a function of time. In incompressible
flow, average density would decrease. However, the rare regions where the density increases affect
the high-order statistical moments of density, making them increase over time. This is a
manifestation of intermittency~\cite{Kraich, Frisch, BalkFoux, ZS}.

So, low-order moments of density decrease, while high-order moments grow; thus, there exists some
distinguished order such that the corresponding moment does neither grow nor decrease. This is the
stochastic integral of motion.  The remarkable feature is that, as we will show below, these orders
turn out to be the same (and equal to the space dimension) for lines and surfaces, and completely
independent of the statistics of the flow. In the case of compressible flow the integral takes more
complicated form, and includes necessarily densities of surfaces of different dimensions embedded
within each other.

In what follows, it is convenient to consider inverse densities, which are norms of differential
forms associated with the evolving surfaces.
 There are no restrictions on the space dimension of the flow, so we will formulate our
results and the proof for arbitrary space dimension $d$ and material surfaces of arbitrary
dimensions $1\le k \le d$ trapped in the flow.

\vspace{1cm}

So,  consider $d$-dimensional ($d \ge 2$) space filled with continuous set of particles (fluid).
The motion of the particles is described by random velocity field $\bu(t,\br)$:
$$
d \br /dt = \bu (t,\br) \ , \ \  \br(0)=\bx
$$
%
We require stochastic isotropy of the vector field $\bu$; this means that
for any rotation $\bR \in SO(d)$ of the space around any point $\ba$,
$$
\br \to \br' = \bR (\br - \ba) + \ba \ , \\
\bu \to \bu'(t, \br') =\bR \bu (t, \br)
$$
and for any functional $F\left[\bu\right]$,  the average of $F$ over all realizations of
$\bu(t,\br)$ does not change under the transformation:  
$$
\langle F\left[\bu' \right]\rangle = \langle F \left[\bu\right]\rangle
$$

Mark all particles by their initial positions $\bx$. Then
\begin{equation}  \label{defQ}
d \br (t,\bx) = \bQ (t,\bx)  d \bx
\end{equation}
where $\bQ$ is the evolution operator; it is also called Jacobian matrix.  
Since $\bQ(t,\bx)$ is a functional of the velocity $\bu$, it
 satisfies the isotropy condition:
$$
\langle F\left[\bR \bQ \bR^T \right]\rangle = \langle F\left[\bQ\right]\rangle \ \  \forall \ \bR\in SO(d)
$$

Now, we consider the initial coordinate grid $\{ x^i \}$ and  the basis vectors $\be_i  $ such that
$\bx = \sum \be_i x^i $; then
$$
\be_i = \frac {\partial\bx}{\partial x^i}
$$
As time goes, the initial coordinate grid undertakes deformations moving with the flow, and the
tangent vectors evolve into
$$
\bl_i (t,\bx) = \frac {\partial\br (t,\bx)}{\partial x^i} \ , \ \  \bl_i (0,\bx) = \be_i
$$
Taking the derivative of (\ref{defQ}) we find that the evolution of $\bl_i$ is described by a
similar equation:
\begin{equation} \label{l-Qe}
 \bl_i (t,\bx) = \bQ(t,\bx) \be_i
\end{equation}
Consider the norms of the differential forms constructed from these vectors:
\begin{equation}  \label{norms}
s_0=1 , \ s_1 = \| \bl_1 \| , \  s_2= \| \bl_1 \wedge \bl_2 \|,..., s_{d} = \| \bl_1 \wedge ...
\wedge \bl_{d} \|
\end{equation}
They describe the evolution of length/area/volume of the $k$-dimensional infinitesimal initial line
segment/square/cube as it evolves and bends in the flow:
$$
d \bsigma^{(k)} = s_k dx_1 \dots dx_k
$$
The differential forms are related to the $k$-dimensional-surface densities by
$$
s_k = \rho_k^{-1}
$$

While geometrically $\left\{ s_k \right\}$ correspond to the norms (hypervolume) of the deformed
$k$-dimensional material elements, algebraically they can be expressed via determinants: because of
(\ref{l-Qe}), every $s_k$ is equal to the square root of the  $k$-order leading principal minor of
the Gram matrix $\Gamma = Q^T Q$:
$$
s_1^2 = \Gamma_{11} \ , \ \ s_2^2= \left| \begin{array}{ll} \Gamma_{11} & \Gamma_{12} \\\Gamma_{21} & \Gamma_{22}
\end{array} \right| \ , \ \  \dots, \ s_d^2=\det \Gamma
$$

The main result of the paper is: \\
In isotropic flows, for any permutation $1..d \to \pi(1)..\pi(d)$, the equality
\begin{equation} \label{result}
\left \langle s_1 ^{\pi(2)-\pi(1)-1} s_2 ^{\pi(3)-\pi(2)-1}...s_{d-1} ^{\pi(d)-\pi(d-1)-1} s_d
^{d-\pi(d)}
 \right \rangle  = 1
\end{equation}
holds at any time, independently of the flow statistics.

More generally, \\
{\sl   Let $m_1,\dots,m_d$ be arbitrary real numbers. Then the function
$$
{\cal K} (m_1,..., m_d) = \left \langle s_1 ^{m_1-m_2-1} \dots s_{d-1} ^{m_{d-1}-m_d-1} s_d ^{m_d+d}
 \right \rangle
$$
is symmetric in the variables $m_1,...,m_d$. }
\\
In other words, we get a family of identities:
\begin{equation} \label{general}
{\cal K} (m_1,\dots, m_d) = {\cal K} \left( m_{\pi(1)}, \dots, m_{\pi(d)}\right)
\end{equation}
%

We start the proof of (\ref{result}) and (\ref{general}) with the following \\
{\bf Lemma. } \ \ Consider an arbitrary symmetric matrix $\bGamma$ with positive leading principal
minors. For arbitrary integer $p: 1\le p <d$, consider the family of rotations $O_p(\phi)$ in the
plane $(x_p,x_{p+1})$:
$$ \bO_p (\phi) =\begin{array}{l} \small
\left(   \begin{array}{lllrll}  1&&&&& \\ & \ddots &&&& \\
&& \cos \phi & -\sin \phi && \\   && \sin \phi & \cos \phi && \\
&&&&\ddots & \\ &&&&& 1  \end{array} \right)
\\
\quad \   1 \ \ \dots \ \ p \quad \  \ \ p+1 \   \dots    d
\end{array}
$$
Let $s_m^2 (\phi)$, $m=1..d$ be the leading principal minors of the matrix $\bGamma_p (\phi)=
\bO_p^T \bGamma \bO_p$. Then for any real $k$,
\begin{equation} \label{Lemma}
\int \limits_0^{2\pi} d\phi {s_p}^{k-1}(\phi) = \int \limits_0^{2\pi}  d\phi \frac {(s_{p-1}
s_{p+1})^{k}}{s_p^{k+1}(\phi)}
\end{equation}

{\bf Proof of the Lemma: }

{\bf 1.} Consider the matrix $\bGamma_p (\phi) = \bO_p^T \bGamma \bO_p$; we are interested in its $p$-rank leading
principal minor. The elements of the matrix 
are linear combinations of the 
elements of~$\bGamma$:
\begin{equation}  \label{8b}
\begin{aligned}
&\left( \bGamma_p(\phi) \right)_{i,j} = &&\bGamma_{i,j}  \ ,&& i,j<p ; \\
&\left( \bGamma_p(\phi) \right)_{i,j} = &&\bGamma_{p,i} \cos \phi +\bGamma_{p+1,i} \sin \phi
 \ ,&& i=p,\, j<p ;  \\
&\left( \bGamma_p(\phi) \right)_{i,j} = &&\bGamma_{p,p} \cos^2 \phi + \bGamma_{p+1,p+1} \sin^2\phi + \\
&  &&\bGamma_{p,p+1} \sin 2\phi \ ,&& i=j=p.
\end{aligned}
\end{equation}
 Since no elements of $\bGamma$ with row or column number $\ge p+2$ are involved in $s_p^2 (\phi)$,
  we temporarily restrict
our consideration to the $(p+1)\times (p+1)$ left-upper part of the matrix $\bGamma$:
$\hat{\bGamma}_{ij}= \bGamma_{ij}$ for $i,j\le p+1$. Let $M_{a,b,c;m,n,k}$ be minors of
$\hat{\bGamma}$ obtained by deleting the rows $a,b,c$ and the columns $m,n,k$.

Making the Laplace decomposition along the $p$-th column and then the same decomposition of the
minors along the $p$-th row, we reduce the minors of $\bGamma_p(\phi) $ to the minors of
$\hat{\bGamma}$:
$$
\begin{array}{ll}
s_p^2 (\phi) &= \sum \limits_{i,j=1}^{p-1} \left( \bGamma_p(\phi)  \right)_{i,p} \left( \bGamma_p(\phi)  \right)_{p,j} (-1)^{i+j} M_{i,p,p+1;j,p,p+1} \\
&+ \left( \bGamma_p(\phi) \right)_{p,p} M_{p,p+1;p,p+1}
\end{array}
$$
Substituting (\ref{8b}) and making use of the same decomposition taken backwards, we get
$$
s_p^2 (\phi) = M_{p+1;p+1} \cos^2 \phi + M_{p;p} \sin^2 \phi + M_{p;p+1} \sin 2\phi
$$
This quadratic (with respect to $\cos \phi,\sin \phi$) form can be diagonalized  by means of an
orthogonal transformation:
\begin{equation} \label{sp2}
s_p^2 (\phi) = u \cos^2 (\phi-\phi_0) + v \sin^2 (\phi-\phi_0)
\end{equation}
where $\phi_0$ is the rotation angle, $u,v, \phi_0$ are independent of $\phi$; for definiteness,
let $u>v$. Since rotation preserves the determinant of the associated matrix,
\begin{equation} \label{uv}
uv = M_{p,p} M_{p+1,p+1} - M_{p,p+1}^2
\end{equation}

Since $M_{ij}$ are first minors of $\hat{\bGamma}$,  (\ref{uv}) is equal to the determinant of the
2x2 right-lower principal minor of the matrix $(\hat{\bGamma})^{-1}$ multiplied by $(\det
\hat{\bGamma})^2$. According to the Frobenius formula  for block matrix inversion \cite{Gantmaher},
it can be expressed in terms of the $p-1$-order leading principal minor of $(\hat{\bGamma})$; it
follows that
\begin{equation}   \label{uv-new}
\begin{array}{ll}
  uv &= M_{p,p} M_{p+1,p+1} - M_{p,p+1}^2
\\
 &= \left( \hat{\bGamma}^{-1}_{p,p}  \hat{\bGamma}^{-1}_{p+1,p+1}-
\left( \hat{\bGamma}^{-1}_{p,p+1}\right)^2 \right) 
(\det \hat{\bGamma})^2
\\ &= s_{p-1}^2 
\det \hat{\bGamma} = s_{p-1}^2 s_{p+1}^2
  \end{array}
\end{equation}
Both $s_{p-1}$ and $s_{p+1}$ are invariant under the rotation ${\bO}_p$, so they do not depend on $\phi$.

{\bf 2.} 
We now make use of (\ref{sp2}).  Making the change of variables
$$
 x= u \cos^2 (\phi-\phi_0) + v \sin^2 (\phi-\phi_0)
$$
in the left-hand side of (\ref{Lemma}), we get
$$
\int \limits_0^{2\pi} d\phi {s_p}^{k-1}(\phi) = 2 \int \limits_v^u \frac{dx
x^{(k-1)/2}}{\sqrt{(u-x)(x-v)}}
$$
Now, changing variables from $x$ to $y=uv/x$ and back to $\phi$, we arrive at
$$
\begin{array}{ll}
\int \limits_0^{2\pi} d\phi {s_p}^{k-1}(\phi) &= (uv)^{k/2} \, 2 \int \limits_v^u \frac{dy}
{y^{(k+1)/2}  \sqrt{(u-y)(y-v)}} \\
&= (uv)^{k/2} \, \int \limits_0^{2\pi} \frac{d\phi}{{s_p}^{k+1}(\phi)} \ ,
\end{array}
$$
With account of  (\ref{uv-new}), this  {\bf proves the Lemma}.

\vspace{1cm}

Now we proceed to the  \\
{\bf Proof of (\ref{result}) and (\ref{general}): } 
\begin{enumerate}
\item
The rotation $O_p$ preserves leading principal minors of the orders $\le p-1$ and $\ge p+1$:
$s_i(\phi)=s_i(0), \ i\ne p$. Thus, from (\ref{Lemma}) we have
\begin{equation} \label{zero}
\int d\phi  \left( s_p^{k-1}(\phi) - s_{p-1}^{k}  s_p^{-k-1}(\phi)
s_{p+1}^{k} \right) \prod\limits_{i\ne p} s_i^{\alpha_{i}} =0
\end{equation}
for any $\alpha_1,...\alpha_d$.

\item From the isotropy condition it follows that
$$
\left \langle \int d\phi \emph{F} \left( \bGamma_p(\phi)  \right) \right \rangle = \int d\phi
\left \langle \emph{F} \left( \bGamma_p(\phi)  \right) \right \rangle = 2\pi \left \langle
\emph{F} (\bGamma) \right \rangle
$$
So, if $\int d\phi \emph{F} \left( \bGamma_p(\phi)  \right)=0$ then $\left \langle \emph{F} \right \rangle=0$; then from
(\ref{zero}) we get    
$$
\left \langle s_1^{\alpha_1}... s_{p-1}^{\alpha_{p-1}} \left( s_p^{k-1} - s_{p-1}^{k}  s_p^{-k-1}
s_{p+1}^{k} \right) s_{p+1}^{\alpha_{p+1}}...s_{d}^{\alpha_d} \right \rangle =0
$$

\item
Taking $\alpha_i=m_i-m_{i+1}-1$,  $i=1..d-1$, $\alpha_d = m_d +d$, $k=m_p-m_{p+1}$ we prove the
special case of (\ref{general}) where only two indices are rearranged: $ \pi(p) = p+1 \ , \quad
\pi(p+1)=p \ , \quad \pi(i\ne p,p+1)=i$.

\item
Any permutation can be obtained by successive  permutations of neighboring elements. So, the
fulfillment of the statement for each pair permutation $p \leftrightarrow p+1$ with $p$ from 1 to $d-1$,
proves the whole statement (\ref{general}).

\item
Take now $m_1=-1, m_2=-2,...,m_d=-d$. Then the left-hand side of (\ref{general}) is identically
equal to one, and we arrive at (\ref{result}).
\end{enumerate}
{\bf End of the proof.}

\vspace{1cm}

The statements (\ref{result}), (\ref{general}) are formulated as averages over the ensemble of
flows $\bu (t,\br)$. However, since we consider homogenous and isotropic flows, in the case of
finite correlation length this result can be presented as the average over the whole volume of the
flow, or as the Lagrangian average over an infinite surface frozen in the flow and travelling with
it. %
To relate the expectation values to measurable quantities, consider a $k$-dimensional infinite
initial surface $S^{(k)}(0)$  and denote  by $S_R^{(k)}(0)$ its transection with a $d$-dimensional
ball of radius $R$. The existence of a finite correlation scale implies that the integrals over
$S_R^{(k)}(0)$ converge
(in probability)
 as $R \to \infty$ to the expectation values:
$$
\lim \limits_{R \to \infty} \frac 1{|S_R^{(k)}(0)|}  \int \limits_{S_R^{(k)}(0)} s_1^{\alpha_1}
\dots s_d^{\alpha_d} d\bx  = \left \langle s_1^{\alpha_1} \dots s_d^{\alpha_d} \right \rangle
$$
This is an analogy of the law of Large numbers.  The integrals over $x$ can be expressed as
integrals over the frozen surface $S_R^{(k)}(t)$ that coincided with $S_R^{(k)}(0)$ at the initial
moment. To this purpose, we note that
%
$d\bsigma^{(k)}=s_k(t,\bx)d\bx$,
so,
$$
 \left \langle s_1^{\alpha_1} \dots s_d^{\alpha_d} \right \rangle =
\lim \limits_{R \to \infty} \frac
1{\scriptstyle{|S_R^{(k)}(0)|}}  \int \limits_{\scriptstyle{S_R^{(k)}(t)}} s_1^{\alpha_1} \dots  s_k^{\alpha_k-1} \dots
s_d^{\alpha_d}   d\bsigma^{k}
$$
%
 Thus, one can represent the flow as a system of hyper-surfaces of different
dimensions embedded in each other and marked at the initial moment by a homogenous layer of paint.
The surfaces undergo deformation as they move with the flow; the density of the paint changes as
$\rho_m = s_m ^{-1} (t,\bx)$ for each $m$-dimensional surface. Then the equations
(\ref{result}),(\ref{general}) can be expressed as the integrals of combinations of surface
densities of different dimensions taken over one hyper-surface trapped in the flow.
 This geometrical interpretation allows to consider the equalities (\ref{result}),(\ref{general})
as integral symmetry properties of any isotropic flow.

In particular, for a cyclic permutation with $\pi(k)=d$, in (\ref{result}) we then get
\begin{equation}\label{cyclic}
\lim \limits_{R\to \infty} \frac{1}{\scriptstyle{|S_R^{(k)}(0)|}} \int \limits_{\scriptstyle{S_R^{(k)}(t)}} \rho_k ^{d+1}
\rho_d^{-k} d\bsigma^{(k)} = \left \langle s_k^{-d} s_d^{k} \right \rangle = 1
\end{equation}
We note that all the exponents are independent of the details of the flow. In incompressible flow,
$\rho_d=const$, and we can restrict ourself by density of only one type.
\begin{equation}\label{A}
\lim \limits_{R\to \infty} \frac{1}{\scriptstyle{|S_R^{(k)}(0)|}} \int \limits_{\scriptstyle{S_R^{(k)}(t)}} \rho_k ^{d+1}
 d\bsigma^{(k)} = \left \langle s_k^{-d} \right \rangle = 1
\end{equation}
This case illustrates
intermittency most clearly: while in most regions $\rho_k$  decreases, it still has to grow in some
regions, to provide constancy of (\ref{A}). For $k=1$ we obtain just the integral of motion found
in \cite{FalkFrishman}. It is noteworthy that the powers of $s_k$ are the same for all integrals of
the set, independently of $k$.

 For compressible case, at least two densities of different
dimensions are involved in every integral of motion. However, for any pair $k<m$ there exists an
integral that contains only two types of densities $\rho_k$ and $\rho_m$:
$$
\lim \limits_{R\to \infty} \frac{1}{\scriptstyle{|S_R^{(m)}(0)|}} \int \limits_{\scriptstyle{S_R^{(m)}(t)}} \rho_k ^{m}
\rho_d^{-k+1} d\bsigma^{(m)} = \left \langle s_k^{-m} s_m^{k} \right \rangle = 1
$$
For the permutation of only two neighbor numbers, $k$ and $k+1$, we have, for instance,
$$
\lim \limits_{R\to \infty}  \frac{1}{\scriptstyle{|S_R^{(k)}(0)|}}  \int \limits_{\scriptstyle{S_R^{(k+1)}(t)}} \rho_{k-1}^{-1}
\rho_k ^{2} d\bsigma^{(k+1)} =  \left \langle s_{k-1} s_k^{-2} s_{k+1} \right \rangle = 1
$$
 If we fix in
(\ref{result}) $\pi(i)=i$ for $i\ge k+1$, we obtain the expression that includes only densities of
smaller dimensions enclosed in the $k$-dimensional hypersurface:
$$
\lim \limits_{R\to \infty} \frac{1}{\scriptstyle{|S_R^{(k)}(0)|}} \int \limits_{S_R^{(k)}(t)} 
\rho_k ^{\pi(k)-k+1}\prod_{i=1}^{k-1}\rho_i ^{\pi(i)-\pi(i+1)+1}
    d\bsigma^{(k)} =1
$$
From Eq. (\ref{result}) we get a total of $d!-1$ integrals of motion, according to the number of
nontrivial permutations.  However, (\ref{general}) produces much more integrals of the form
$$
{\cal K}(m_1,\dots,m_d)/{\cal K} ( m_{\pi(1)},\dots,m_{\pi(d)}) =1
$$
For example, for the permutation of two neighbor elements we get
$$
{\left \langle s_1^{\alpha_1}\dots s_d^{\alpha_d} s_{k-1}^{\alpha_k+1} s_k^{-\alpha_k-1}
s_{k+1}^{\alpha_k +1} \right \rangle \over \left \langle s_1^{\alpha_1}\dots s_d^{\alpha_d}
 \right \rangle } =1
$$

All these relations can be applied, e.g., to boundaries between chemical or biological fractions in
a turbulent flow (if diffusion is negligible).

We note that the existence of  all the integrals of motion found in the paper requires three
assumptions: statistical isotropy of the flow, smooth velocity field,  and surfaces frozen in the
flow. Thus, all the conclusions remain valid also if the structures and quantities under
consideration may dynamically influence the flow. These are, for example, magnetic field in highly
conductive liquid, or vorticity and vortex lines  in Eulerian turbulence. In these cases, the
absolute values of magnetic induction $B$ and vorticity $\omega$ play the role of $s_1$. From
(\ref{cyclic}) it follows that the statistical moment  $\langle
\omega^{-3}\rangle$ (or $\langle \omega^{-3} V\rangle$ for compressible isentropic flow) is
preserved if viscosity does not affect the vorticity dynamics.
Even if viscosity is significant, $\langle\omega^{-3}\rangle$ can still conserve approximately, since viscosity manifests itself in areas where vorticity is high, while
the moment is mainly contributed by the regions of low vorticity.
 Similarly, $\langle B^{-3}\rangle$ (or $\langle B^{-3}
V\rangle$, for compressible flow) is conserved as long as one can neglect magnetic diffusivity,
i.e., as long as there is no reconnection of magnetic lines. If the magnetic feedback stops the
flow before magnetic diffusivity comes into play, the relations still remain valid.

One more comment concerns the ergodic theory. The result obtained in the paper makes some specific
restrictions on the symmetry of the Kramer function. Specifically, the probability density formulated in terms of the forms
$s_k$
%
 must satisfy the requirement of invariance of ${\cal P}(s_1,...,s_p,...,s_d)$
under the transformation $s_p \to s_{p+1} s_{p-1}/s_p$.
This may be helpful in constructing the explicit form of the Kramer function in different
models~\cite{fin-time, JOSS1, LagrModel}.

The authors are thankful to Professor G. Falkovich for drawing our attention to the problem. We also thank the Anonymous Referee for his valuable remarks.
This work of A.V. Kopyev was supported by the Foundation for the Advancement of Theoretical Physics and Mathematics (BASIS) 23-1-3-46-1.
The data that supports the findings of this study are available within the article.

\end{document}